\documentclass[apj, iop]{emulateapj}
\usepackage{apjfonts}

\usepackage{natbib}
\usepackage{ulem}

\newcommand{\MO}{$M_\odot$}

\newcommand{\Suzaku}{\textit{Suzaku}}
\newcommand{\Chandra}{\textit{Chandra}}
\newcommand{\Swift}{\textit{Swift}}
\newcommand{\NH}{$N_{\rm H}$}

\shorttitle{\Suzaku\, observations of the type 2 QSO of the Phoenix cluster}
\shortauthors{Ueda et al.}

\begin{document}


\title{\textit{Suzaku} observations of the type 2 QSO in the central galaxy of the Phoenix cluster}


\author{Shutaro~Ueda\altaffilmark{1}, Kiyoshi~Hayashida\altaffilmark{1}, Naohisa~Anabuki\altaffilmark{1},
Hiroshi~Nakajima\altaffilmark{1}, Katsuji~Koyama\altaffilmark{1,2}, and Hiroshi~Tsunemi\altaffilmark{1}}
\altaffiltext{1}{Department of Earth and Space Science, Graduate School of Science, Osaka University, 1-1, Machikaneyama, Toyonaka, Osaka, 560-0043, Japan}
\altaffiltext{2}{Department of Physics, Graduate School of Science, Kyoto University, Kita-Shirakawa, Sakyo-ku, Kyoto, 606-8502, Japan}
\email{shutaro@ess.sci.osaka-u.ac.jp}

\begin{abstract}
We report the \Suzaku/XIS \& HXD and \Chandra/ACIS-I results on the X-ray spectra of the Phoenix cluster at the redshift $z=0.596$. 
The spectrum of the intracluster medium (ICM) is well-reproduced with the emissions from
 a low temperature ($\sim3.0$\,keV and $\sim0.76$\,solar) and a high temperature ($\sim11$\,keV and $\sim0.33$\,solar) plasmas; 
the former is localized at the cluster core, while the latter distributes over the cluster. 
In addition to these ICM emissions, a strongly absorbed power-law component is found, which is due to an active galactic nucleus (AGN) in the cluster center. 
The absorption column density and unobscured luminosity of the AGN are $\sim3.2\times10^{23}$\,cm$^{-2}$ and $\sim4.7\times10^{45}$\,ergs\,s$^{-1}$ ($2-10$\,keV), respectively. 
Furthermore, a neutral iron (\ion{Fe}{1}) K-shell line is discovered for the first time with the equivalent width (EW) of $\sim150$\,eV at the rest frame. 
The column density and the EW of the \ion{Fe}{1} line are exceptionally large for such a high luminosity AGN, and hence the AGN is classified as a type 2 quasi-stellar object (QSO). 
We speculate that the significant fraction of the ICM cooled gas would be consumed to maintain the torus and to activate the type 2 QSO. 
The Phoenix cluster has a massive starburst in the central galaxy, indicating suppression in the cooling flow is less effective. 
This may be because the onset of the latest AGN feedback has occurred recently and it has not yet been effective. 
Alternatively, the AGN feedback is predominantly in radiative-mode not in kinetic-mode and the torus may work as a shield to reduce its effect.
\end{abstract}



\section{Introduction}

Brightest cluster galaxies (BCGs) are giant elliptical galaxies located at the centers of the clusters.  
BCGs host supermassive black holes (SMBHs) at their nuclei as for other massive galaxies.  
BCGs and SMBHs in them are located at the highest end of the well known $M_{\rm BH}$ and $\sigma$ relation, 
where $M_{\rm BH}$ is the mass of a SMBH and $\sigma$ is the velocity dispersion of the bulge of a galaxy \citep[e.g.][]{Ferrarese00, Gebhardt00, Tremaine02, Sadoun12, Salviander13, McConnell13}. 
However, a large scatter or deviation from the $M_{\rm BH}$ and $\sigma$ relation extrapolated from the non BCG sample is found for the SMBHs in BCGs. 
Observational evidence and numerical simulations indicate that highest mass SMBHs ($\sim 10^{10}$\MO) and 
their host BCGs undergo some unique history of merging and/or accretion processes \citep[e.g.][]{Hopkins07, Gultekin09, McConnell11, McConnell12, Graham13, Volonteri13}.

The number density of distant quasi-stellar objects (QSOs) is larger than that of nearby QSOs \citep{Richards06},
while \cite{McConnell11} found that the number density of nearby BCGs is consistent with that of SMBHs in the highest mass limit ($10^{9} - 10^{10}$\,\MO),
which are predicted from the $M_{\rm BH}$-$L$ relation \citep[e.g.][]{Ferrarese00, McLure02, Marconi03} and the luminosity function of nearby galaxies.
\cite{McConnell11} suggests that local BCGs host the remnants of highly luminous QSOs.
Some numerical simulations indicate that luminous QSOs in the high-redshift end up as QSOs 
in the massive central galaxies (i.e. BCGs) of rich clusters at the local Universe \citep[e.g.][]{Springel05, Springel06, Li07, Angulo12}.
Observationally, \cite{De_Lucia07} shows that high-redshift BCGs belong to the same populations of local BCGs in their massive end, 
while \cite{Husband13} indicates that the luminous QSOs at $z \sim 5$ likely represent an early stage in building-up massive low-redshift clusters.
QSOs would be very active in the early Universe, possibly at $z = 2 - 4.5$, but are dormant at present \citep[e.g.][]{Richards06, Vestergaard08}.  
For example, the prominent radio-jet galaxy M87 in the Virgo cluster hosts a SMBH of $6.3 \times 10^{9}$\,\MO\, \citep{McConnell11},  
but the present luminosity is many orders of magnitude lower than the Eddington limit \citep{Di_Matteo03b}, much lower than those of typical active galactic nuclei (AGNs). 

Detailed study of the active phase of the SMBHs in BCGs is difficult for the sources at $z>1$, 
but there are exceptional cases in which QSOs are found in the BCGs of clusters at lower redshift. 
In such sources, SMBHs are surely growing by gas accretion \citep[e.g.][]{Salpeter64, Hopkins05b}, 
while strong radiation and/or jets from the SMBHs might affect the intracluster medium (ICM) in clusters \citep[e.g.][]{Fabian12}. 
Therefore, those sources are of extreme importance for the study of the feeding and feedback processes in SMBHs, BCGs, and clusters.  

Several such candidates include E1821+643 \citep{Kii91, Yamashita97, Jimenez_Bailon07, Russell10}, 
3C\,186 \citep{Siemiginowska10}, PKS\,1229-021 \citep{Russell12}, IRAS\,09104+4109 \citep{Iwasawa01, Vignali11, OSullivan12}, IRAS\,F15307+3252 \citep{Iwasawa05},
and the Phoenix cluster \citep{McDonald12, McDonald13}. 
Among them, we select the Phoenix cluster at the redshift $z = 0.596$, hosting a massive BCG with very luminous SMBH.

An X-ray emission from the Phoenix cluster is firstly reported as 1RXS\,J234444.1-424319 in the \textit{ROSAT} Bright Source Catalog \citep{Voges99}. 
This source is classified to be a Seyfert 2 in the Quasar and AGN Catalog 10th Edition by \cite{Veron-Cetty01}.  
The Two Micron All Sky Survey (2MASS) found an extended source 2MASX\,J23444387-423124 \citep{Skrutskie06}, 
while the Palermo \Swift/BAT hard X-ray catalog source, 2PSBC\,J2344.8-4245 is identified as 2MASX\,J23444387-423124  \citep{Cusumano10}. 
The $14-150$\,keV band luminosity is extremely high as $1.4 \pm 0.9 \times 10^{46}$\,ergs\,s$^{-1}$.

First identification of this source as a cluster was, however, made with the South Pole Telescope (SPT) via the Sunyaev-Zel'dovich effect, and named as SPT-CLJ2344-4243 \citep{Williamson11}.
With radio, infrared, optical, ultraviolet, and X-ray observations of this source, 
\cite{McDonald12} reported that the X-ray luminosity within $r_{500}$
and the total mass within $r_{200}$ of this cluster are $8.2 \times 10^{45}$\,ergs\,s$^{-1}$ in $2-10$\,keV
and $2.5 \times 10^{15}$\,\MO, respectively.
Both are exceptionally large compared to the other known clusters.

Using \textit{Hubble Space Telescope} and \Chandra \,data \citep[the same data as][]{McDonald12}, \cite{McDonald13} 
estimated that the SFR and cooling rate are $798 \pm 42$\,\MO \,yr$^{-1}$ and $2700 \pm 700$\,\MO \,yr$^{-1}$ \citep[also from][]{White97}, respectively.
Thus, the SFR is $30 \pm 8$\,\% of the cooling rate,
which is one of the highest among typical cool-core clusters in the local Universe.
\cite{McDonald12} also reported the central AGN has the luminosity of $\sim 3 \times 10^{45}$\,ergs\,s$^{-1}$ ($2-10$\,keV) with a large absorption of $\sim 3.9 \times 10^{23}$\,cm$^{-2}$.  

Apart from these general features of the Phoenix cluster, no detailed X-ray spectroscopy
especially for the central AGN, and hence no accurate physical parameters of accretion and obscuration from X-ray observation \citep[e.g.][]{Mushotzky93},
has been examined so far, probably due to the limited statistics in the previous observations.  
Furthermore, the \Swift/BAT hard X-ray, a key band to evaluate the intrinsic luminosity of the AGN, would be contaminated by the thermal plasma in the cluster depending on the plasma temperature.  
We therefore examined the high quality data of the deep \Suzaku\, observation. 
To separately examine the central AGN component from the thermal emission of the ICM, we also employed the \Chandra\, archival data. 

In this paper, we adopt the abundance table of \cite{Anders89}, the Hubble constant of $H_{0} = 70$\,km\,s$^{-1}$\,Mpc$^{-1}$, $\Omega_{\rm M} = 0.27$, and $\Omega_{\Lambda} = 0.73$. 
One arcsec corresponds to 6.7\,kpc at the redshift $z=0.596$ for this cluster. 
Unless otherwise specified, all errors represent at 90\,\% confidence level (90\,\% CL).

\section{Observations and Data Reductions}

A \Suzaku \, \citep{Mitsuda07} observation of the Phoenix cluster was performed on Nov.\,15th-16th 2010 (ObsID:70549010, PI: W. Baumgartner). 
Data process and reduction were done with the HEASOFT version 6.12 (e.g. \texttt{xispi}, \texttt{xselect}).
We first reprocessed the unfiltered event files in the \Suzaku \,data archive and the calibration data base (CALDB) released on Oct.\,15th 2012. 
Other information was derived from the \Suzaku \,team \footnote{http://heasarc.gsfc.nasa.gov/docs/suzaku/analysis/abc/}.
After the standard data reduction, 
the exposure times were 62\,ks and 47\,ks for the X-ray Imaging Spectrometer \citep[XIS:][]{Koyama07} and the Hard X-ray Detector \citep[HXD:][]{Takahashi07, Kokubun07}, respectively. 

\begin{figure*}[ht] 
\begin{center}
\includegraphics[width=66mm, height=55mm]{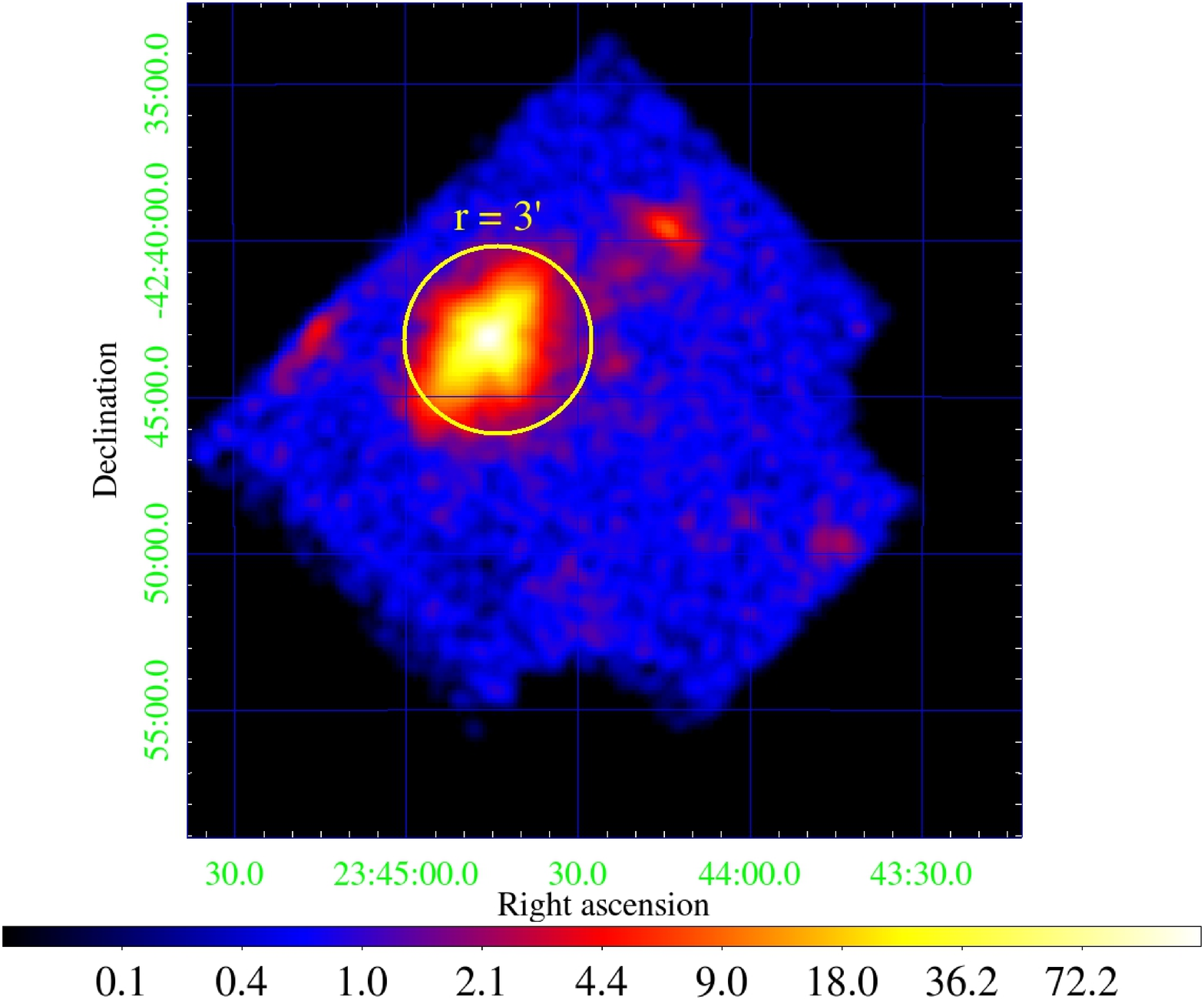}
\includegraphics[width=60mm, height=55mm]{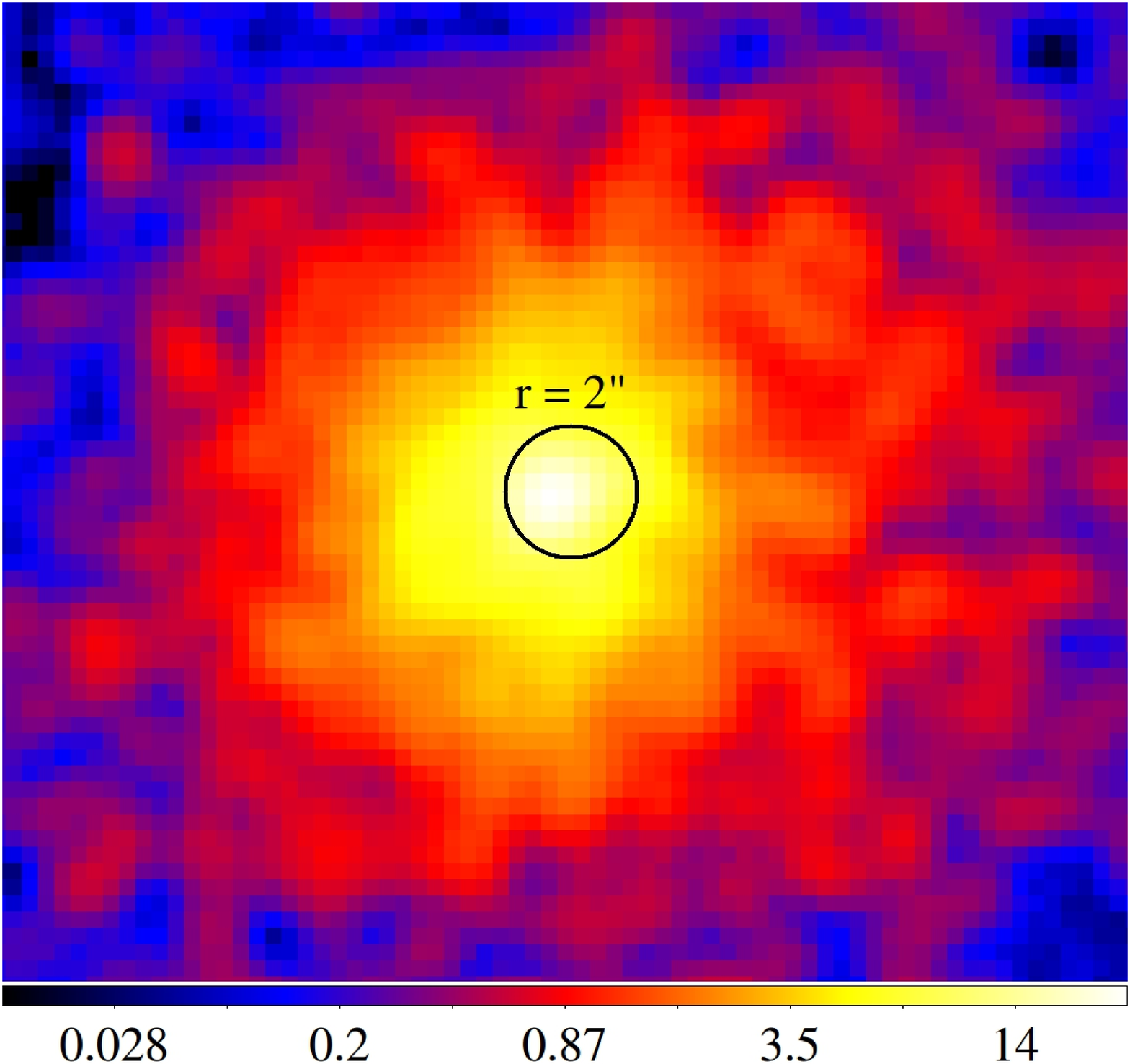}
\caption{
X-ray images of the Phoenix cluster in the $0.4-10$\,keV band. 
The background is not subtracted and the vignetting effect is not corrected. The unit of color bar is counts\,pixel$^{-1}$.
Left panel: X-ray image of \Suzaku/XIS3. 
The yellow circle shows the region of $r<3\arcmin$.
Right panel: The \Chandra/ACIS-I image of central region of the Phoenix cluster. 
The image is smoothed with a 2-dimensional Gaussian of $\sigma = 3$\,pixels. 
The black line is the $r=2\arcsec$ circle.
}
\label{fig:image}
\end{center}
\end{figure*}

The XIS (XIS0, XIS1, and XIS3) data were extracted from the circular region with a radius of $3\arcmin$ centered on the BCG of the Phoenix cluster as shown in the left panel of Figure~\ref{fig:image}. 
The non X-ray background (NXB) of the XIS was estimated  by using the database of night earth observations with \texttt{xisnxbgen} \citep{Tawa08}. 
For the extracted data, we made the NXB-subtracted light curve of XIS0, 1, and 3 in the $0.4-10$\,keV band from the $r<3\arcmin$ region with the time bin size of 1024\,s.  
All the data points are within the range of $\pm 3\,\sigma$ ($\pm0.10$\,cts\,s$^{-1}$) of the mean (1.08\,cts\,s$^{-1}$), 
and hence no significant variability (no anomaly) in the data is found, or no flickering event is included. 

We also used the \Chandra\, archival data of this source (ObsID:13401, PI: G. Garmire), which are the same data used by \cite{McDonald12}. 
The observation was carried out on Sep.\,19th 2011 for an exposure time of 12\,ks with the Advanced CCD Imaging Spectrometer \citep[ACIS:][]{Garmire03}. 
We reprocessed and reduced the level 2 \Chandra/ACIS-I event data by using the \Chandra\, Interactive Analysis of Observations (CIAO) version of 4.4.1 and the CALDB version 4.5.3.

\section{Spectral analyses and results}
\label{sec:fit}

In the analysis of the \Suzaku\, spectra, the NXB is subtracted by using the database (for the XIS) or model (for the HXD), respectively.  
The cosmic X-ray background (CXB) is given by the cut-off power-law \citep{Boldt87}. 
The Milky Way halo (MWH) and the local hot bubble (LHB) are described by the APEC models \citep{Smith01} of 0.23\,keV and 0.07\,keV, respectively \citep{Ueda13b}. 
In the spectral fitting,  
these  X-ray background models with the Galactic absorption \citep[wabs model,][]{Morrison83} of the column density (\NH) $1.52 \times 10^{20}$\,cm$^{-2}$ \citep{Kalberla05} are added to the source model. 

The Redistribution Matrix Files (RMFs) and Ancillary Response Files (ARFs) of the XIS are generated with \texttt{xisrmfgen} and \texttt{xissimarfgen} \citep{Ishisaki07}, respectively. 

The HXD consists of two detectors, PIN and GSO, but we used only the PIN data.  
The RMF (\texttt{ae\_hxd\_pinhxnome9\_20100731.rsp}) and the NXB are released by the HXD calibration team.
The cross-calibration error in the effective areas of different detectors (XIS and HXD) is compensated by multiplying 1.181 to the normalization of the HXD 
(\Suzaku\,Memo\footnote{ftp://legacy.gsfc.nasa.gov/suzaku/doc/xrt/suzakumemo-2008-06.pdf}).
The normalization of the CXB for the HXD is fixed at the values supplied by the HXD team. 

For the \Chandra/ACIS-I spectra, on the other hand, 
we subtract the NXB plus the X-ray backgrounds which were made from the region beyond $7\arcmin$ from the center in the same field of the Phoenix cluster. 
The RMFs and the ARFs are generated by using \texttt{specextract}. 

\subsection{\Suzaku/XIS \& HXD spectra in the $r < 3\arcmin$ region}
\label{sec:r=3}
  
We show the \Suzaku/XIS  image of the Phoenix cluster field in the left panel of Figure~\ref{fig:image}.  
The spectra are extracted from the $r < 3\arcmin$ region (the solid circle in the left panel of Figure~\ref{fig:image}); 
those from XIS0 and XIS3 are restricted to the $0.4-10$\,keV range, while the XIS1 spectrum is restricted to $0.4 - 7$\,keV. 
In the rest frame of $z=0.596$,  $3\arcmin$ corresponds to 1.2\,Mpc, and hence the $r<3\arcmin$ circle includes the major fraction of the cluster emission.  

Since the \Suzaku/HXD is a non-imaging instrument, the spectral data are from the PIN field of $34\arcmin \times 34\arcmin$.  
We employ the energy range of $16 - 40$\,keV. 
The HXD count rate from the source (the NXB and the CXB subtracted) is $1.5 \pm 0.2 \times 10^{-2}$\,cts\,s$^{-1}$ in $16-40$\,keV, 
which corresponds to 7.3\,\% of the NXB, significantly larger than the systematic error in the NXB of $2.1 - 2.7$\,\% ($1\sigma$) \citep{HTakahashi10}.

In addition to the X-ray background model (see section 3), we apply a single temperature (1T) thin thermal plasma model (APEC) as the spectrum of the Phoenix cluster. 
Then, we obtain the gas temperature, abundances, and redshift of the Phoenix cluster, as $16.7 ^{+0.9} _{-0.8}$\,keV, $0.81 \pm 0.09$\,solar, and $0.656 ^{+0.007} _{-0.005}$, respectively.
This 1T model, however, shows the over-all spectral shape with the concave residual, in excess at the low and high energy bands (see the top left panel of Figure~\ref{fig:XIS_HXD}).  
Furthermore, the best-fit redshift $z=0.656$ is inconsistent with the optical observations \citep{McDonald12}.  
We also find a line-like residual at 4.3\,keV, which corresponds to $\sim 7$\,keV after correcting the redshift of 0.656.  
Thus, the residual may be either due to misidentification of the K-shell lines, and hence gave a larger plasma temperature as 16.7\,keV, or due to an additional iron K-shell line. 

The excess at the low energy band may indicate the presence of another thin thermal plasma component with a low temperature.  
Two-temperature structure is already suggested in the 1T fitting of the spatially resolved  \Chandra/ACIS-I spectra \citep{McDonald12}, 
which shows a low temperature in the inner region of $r \lesssim 100$\,kpc, and high temperature in the outer region.  
The excess at the high energy band suggests the presence of a power-law component at the cluster center.

\subsection{\Chandra/ACIS-I spectra from the inner and outer regions}
\label{sec:2-3}

We show the \Chandra/ACIS-I image to highlight the core and surrounding envelope of the Phoenix cluster in the right panel of Figure~\ref{fig:image}.  
Then, we extract the \Chandra/ACIS-I spectra from the inner region of $r<2\arcsec$, and the outer region of $2\arcsec < r < 3\arcmin$ annulus.  

To examine the high temperature plasma in the outer region, we fit the spectrum with an APEC model fixing the redshift to $z = 0.596$.  
This model gives a nice fit with $\chi ^{2}/{\rm d.o.f.} = 89/122$, as is shown in the top right panel of Figure~\ref{fig:XIS_HXD}. 
The best-fit temperature and abundance are constrained to be $10.9 ^{+1.8} _{-1.1}$\,keV and  $0.33 ^{+0.18}_{-0.16}$\,solar, respectively.  
They are roughly consistent with those obtained by \cite{McDonald12}.

We then examine the spectrum from the inner region of $r<2\arcsec$.  
As is shown in the bottom left panel of Figure~\ref{fig:XIS_HXD}, this spectrum has a local minimum at about 2\,keV, which indicates the presence of a soft component plus highly absorbed hard component. 
The former would be a low temperature plasma, while the latter is likely an AGN.
We therefore fit the spectrum with an APEC model (for low temperature plasma) plus an absorbed power-law (absPL) continuum (for AGN).  
This model is accepted  with $\chi ^{2}/{\rm d.o.f.} = 45/42$.  
We thus conclude that the X-rays from the inner region of the Phoenix cluster consist of a low temperature component and an AGN power-law component.  
However, the physical parameters are only loosely constrained.  
The best-fit  photon index and absorption column density for the power-law component are 
$\Gamma = 0.71 ^{+0.82} _{-0.66}$ and \NH $ = 1.9 ^{+2.4} _{-1.2} \times 10^{23}$\,cm$^{-2}$, respectively, 
while the temperature and abundance for the low temperature plasma are $4.89 ^{+6.07} _{-1.80}$\,keV and $1.48 ^{+2.26} _{-1.13}$\,solar, respectively. 
Table~\ref{tab:fit} summarizes the best-fit values of these fittings.

\subsection{\Suzaku/XIS fit in the 3.5-5.0\,keV}
\label{sec:iron}

\begin{figure*} 
\begin{center}
\epsscale{1.0}
\includegraphics[width=72mm, height=54mm]{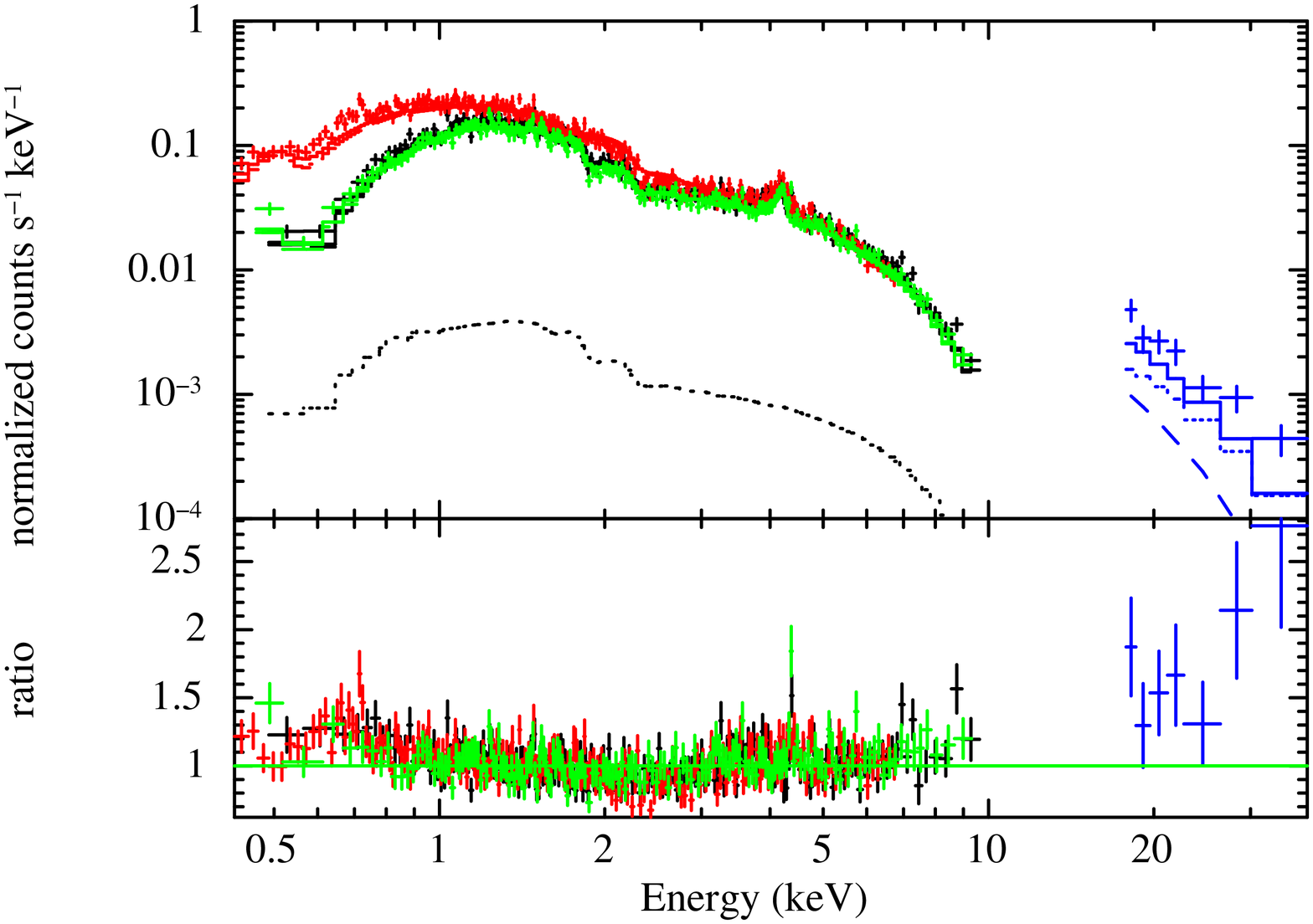}
\includegraphics[width=72mm, height=54mm]{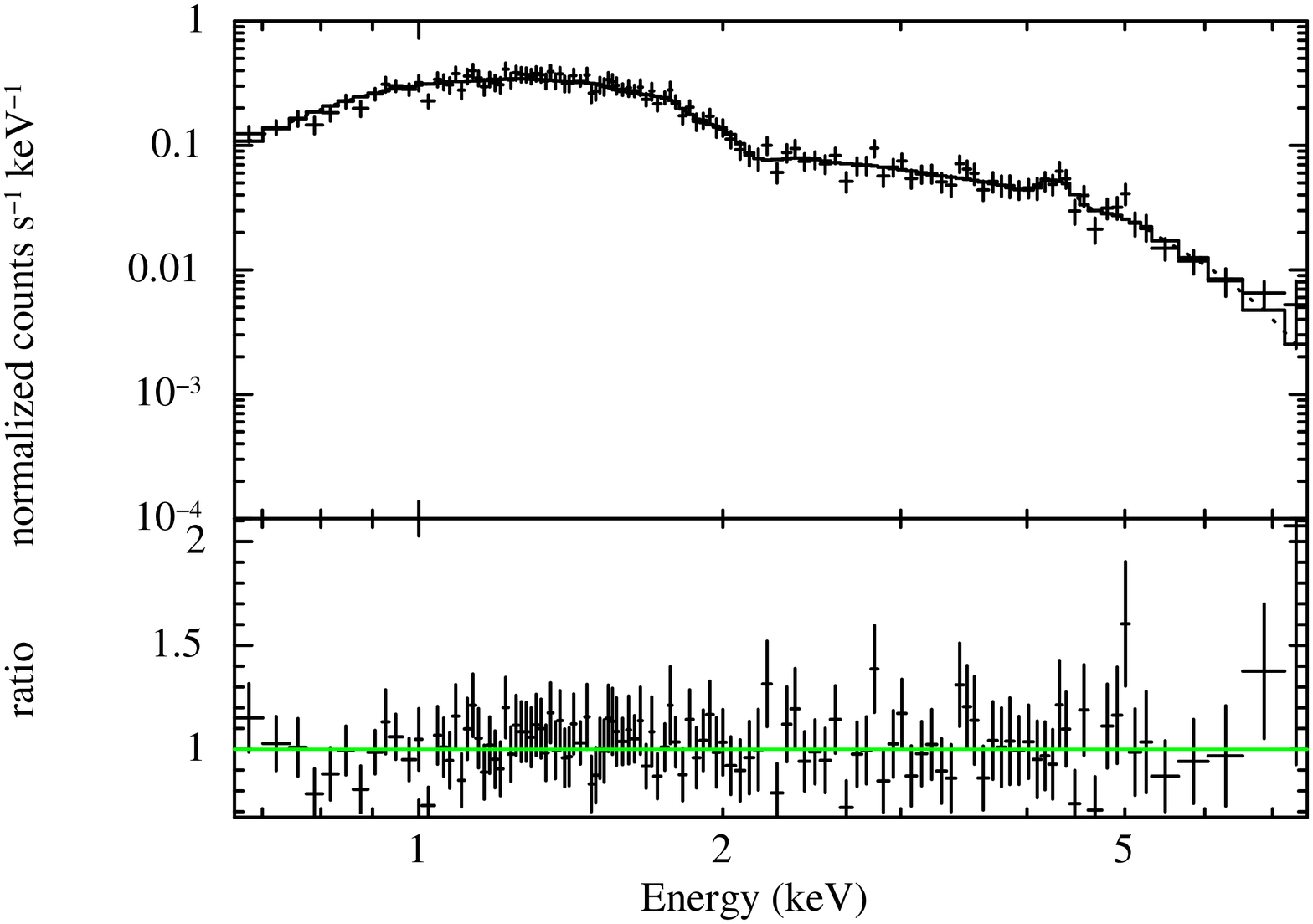}
\\
\includegraphics[width=72mm, height=54mm]{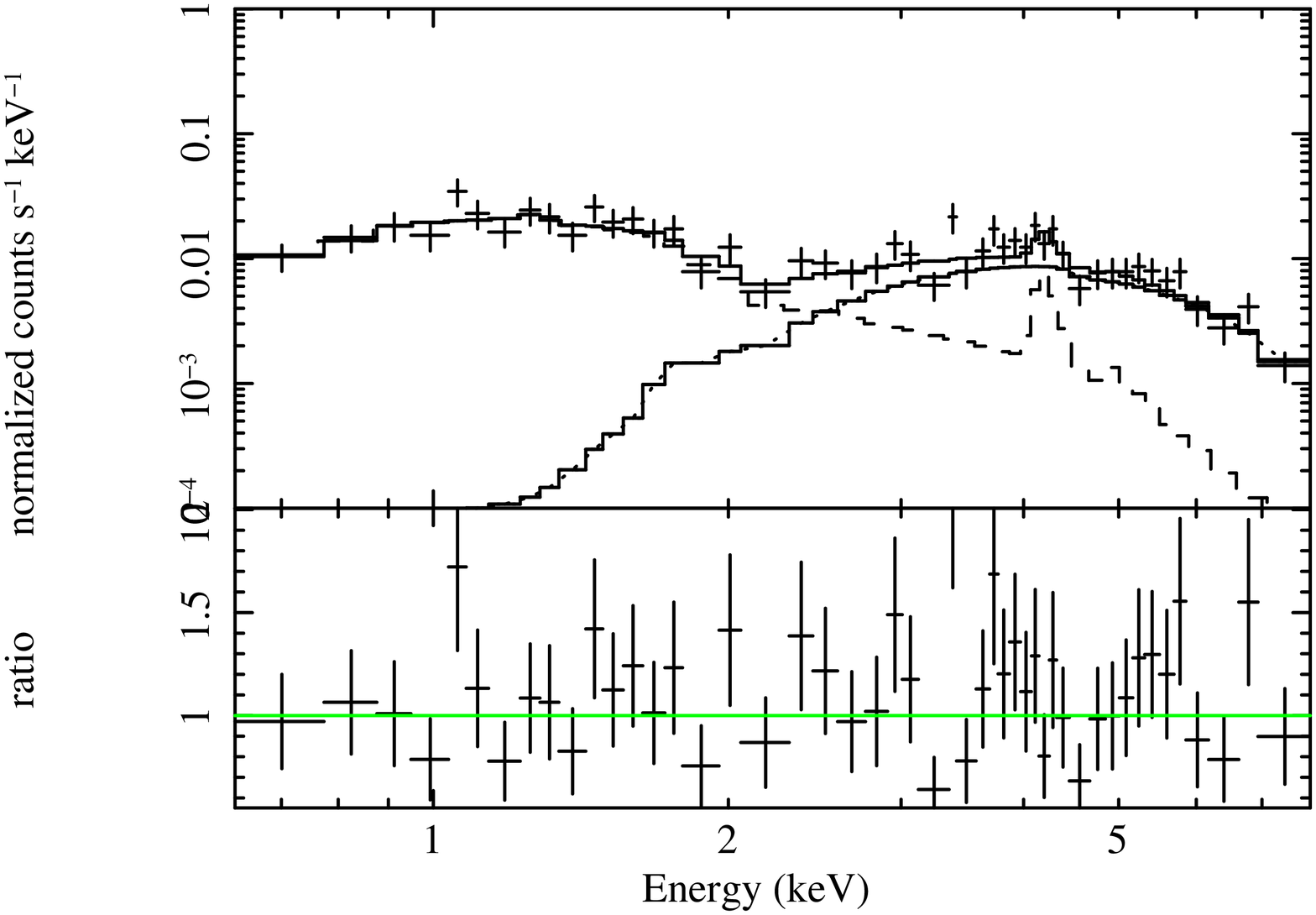}
\includegraphics[width=72mm, height=54mm]{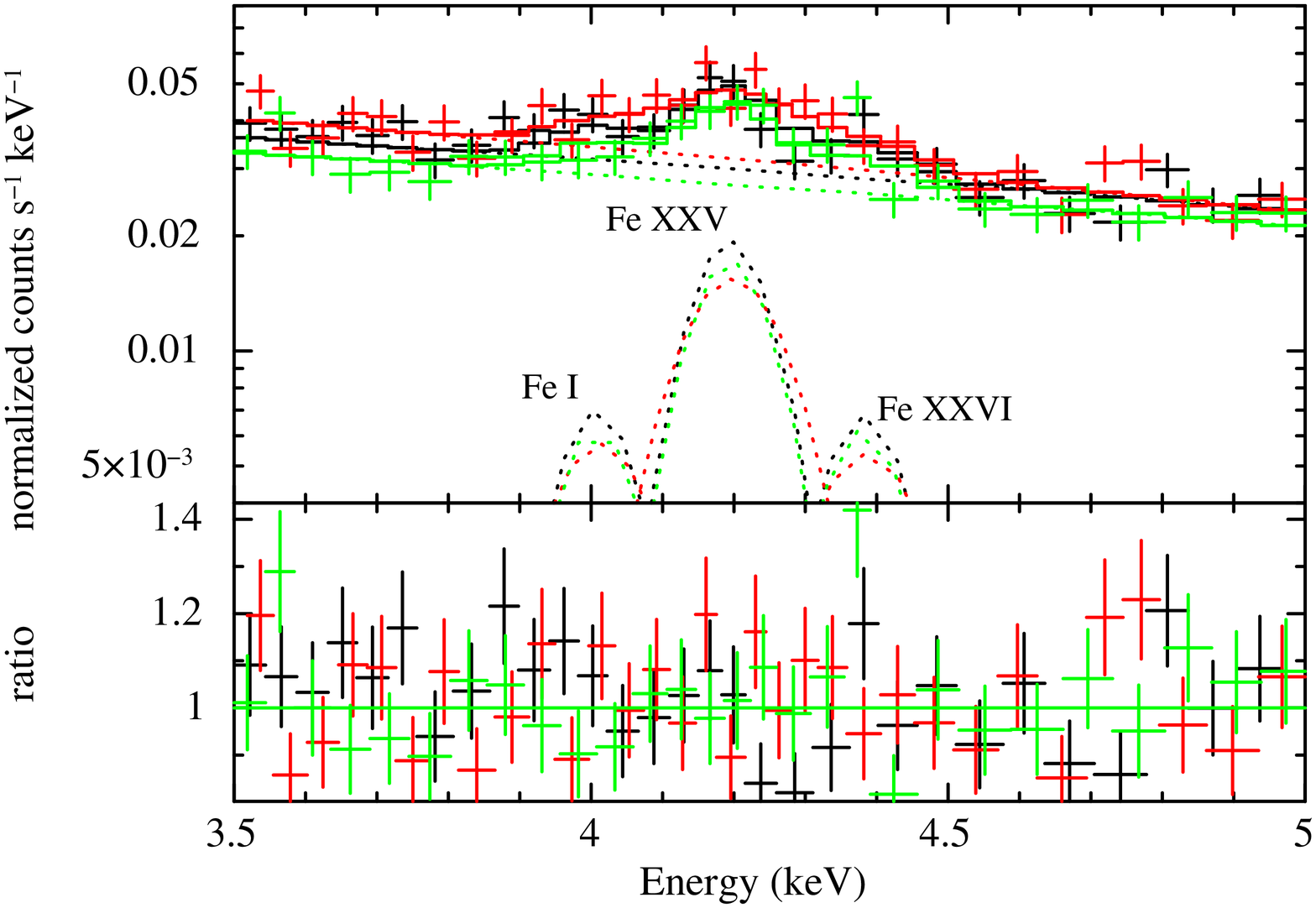}
\caption{
X-ray spectra of the Phoenix cluster. 
The ratios of the data to the model are plotted in the bottom panels.
Top left panel: \Suzaku/XIS0 (black), XIS1 (red), XIS3 (green), HXD (blue) spectra in the $r<3\arcmin$ regions centered at the BCG fitted with a 1T model.
The dashed lines show the best-fit model of the ICM for XIS0 and the HXD.
The dotted lines represent the X-ray background model consisting of CXB, MWH, and LHB for XIS0 and HXD.
Top right panel: \Chandra/ACIS-I spectrum fitted with a 1T model. 
The region is $2\arcsec < r < 3\arcmin$.  
Bottom left panel: \Chandra/ACIS-I spectrum in the core region ($r<2\arcsec$) fitted with a 1T model and an absorption power-law model.
Bottom right panel: \Suzaku/XIS spectra in 3.5$-$5.0\,keV fitted with a power-law continuum and three Gaussian lines. 
}
\label{fig:XIS_HXD}
\end{center}
\end{figure*}

In the 1T model fit for the \Suzaku \,spectra (subsection \ref{sec:r=3}), we also found a significant line-like residual at the energy at $\sim$4\,keV.  
Converting it to the rest frame, the line energy corresponds to either the  K-shell lines from neutral iron (\ion{Fe}{1}), He-like \ion{Fe}{25} or H-like \ion{Fe}{26}. 
We therefore zoom-up the XIS spectrum in the $3.5-5.0$\,keV range, 
and fit with a power-law continuum plus three Gaussian lines, in which the line energies are fixed to those of the K-shell transition from \ion{Fe}{1}, \ion{Fe}{25}, and \ion{Fe}{26} at the redshift $z=0.596$. 
The spectrum and the best-fit results are shown in the bottom right panel of Figure~\ref{fig:XIS_HXD}.
The \ion{Fe}{1} line is detected at 3.5\,$\sigma$ level. 
The equivalent width (EW) defined to the summed continuum of the thermal (i.e. the ICM emissions) and non-thermal (i.e. the central AGN emission) components is $23^{+10}_{-11}$\,eV
at the observer frame. 
Since a thin thermal plasma cannot emit the \ion{Fe}{1} line, the most likely origin is an AGN in the BCG.   

As we suggest in subsection~\ref{sec:r=3}, we find that the 1T model fit of the wide band spectra misidentified the \ion{Fe}{25} line to that of \ion{Fe}{26} and \ion{Fe}{1} to \ion{Fe}{25},  
and hence misled to a larger redshift of 0.656 and higher temperature of $kT \sim 17$\,keV. 

\subsection{Simultaneous fit for the \Suzaku/XIS \& HXD and \Chandra/ACIS-I spectra 
in the $r < 3\arcmin$ region} \label{simfit:r=3}

\begin{figure} 
\begin{center}
\includegraphics[width=72mm, height=54mm]{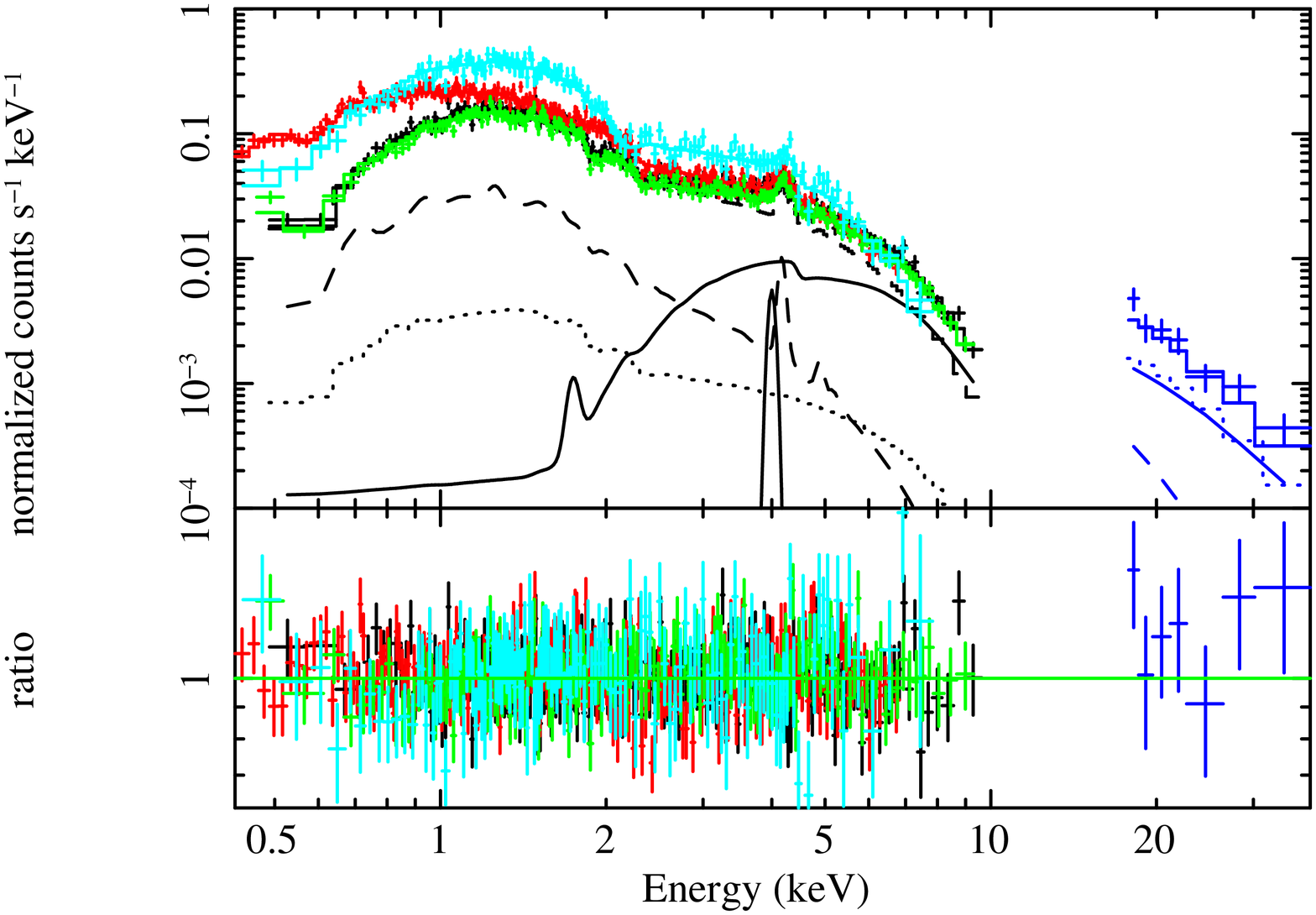}
\caption{
\Suzaku/XIS \& HXD and \Chandra/ACIS-I spectra simultaneously fitted with a 2T$+$absPL$+$\ion{Fe}{1} model expressing the cluster thermal emission and the central AGN emission. 
The same colors and lines are used as those in Figure~\ref{fig:XIS_HXD} for the \Suzaku/XIS0, XIS1. XIS3, and HXD data.
The \Chandra/ACIS-I data is plotted in cyan.
The solid lines show an absorbed power-law component and a neutral iron K-line (\ion{Fe}{1}) for XIS0 and HXD.
}
\label{fig:Suzaku+Chandra}
\end{center}
\end{figure}

We finally carry out the simultaneous fit for the X-ray spectra extracted in the same $r < 3\arcmin$ region from the \Suzaku/XIS \& HXD and \Chandra/ACIS-I.
The model includes all the components found in the previous subsections.
Schematically the spectral model is given as wabs $\times$ (APEC$_{\rm low}$ $+$ APEC$_{\rm high}$ $+$ zwabs $\times$ power-law  $+$ zgauss). 
We fix the temperature and abundance of APEC$_{\rm high}$ component as those obtained in the spectral fit with the \Chandra/ACIS-I spectrum of the $2\arcsec < r < 3\arcmin$ region.  
The line center energy was fixed at 6.40\,keV (i.e. the \ion{Fe}{1} line) at the rest frame. 
Since other parameters determined in the previous subsections are not well constrained, we treat them as free parameters.
This model (2T$+$absPL$+$\ion{Fe}{1}) nicely reproduces the over-all spectra with $\chi ^{2}/{\rm d.o.f.} =790/732$.  
The best-fit model and the data residual are shown in Figure~\ref{fig:Suzaku+Chandra}, while the best-fit parameters are summarized in Table~\ref{tab:fit}.
All the best-fit parameters are consistent with those determined by the individual spectral fit given in subsections \ref{sec:2-3} and \ref{sec:iron}, but are more accurately determined.
For comparisons with the previous results, 
we calculate the $10-50$\,keV and $14-150$\,keV band luminosities for the power-law component as 
$9.4^{+0.1}_{-0.2} \times 10^{45}$\,ergs\,s$^{-1}$ and $2.1 ^{+0.7}_{-0.8} \times 10^{46}$\,ergs\,s$^{-1}$, respectively.  
The EW of the \ion{Fe}{1} line defined to the continuum emission of the central AGN (i.e. the absorbed power-law component) is $149 ^{+139} _{-58}$\,eV  at the rest frame.

\begin{table*}
\begin{center}
\caption{
Best-fit results of the spectral fitting with the \Chandra/ACIS-I in $0.6 - 8$\,keV and the \Suzaku/XIS \& HXD and the \Chandra/ACIS-I in $0.4-40$\,keV.
}\label{tab:fit}
\begin{tabular}{lllclclcc}
\tableline\tableline
Instruments			& Region   & \multicolumn{2}{c}{APEC$_{\rm low}$} & \multicolumn{2}{c}{APEC$_{\rm high}$} & \multicolumn{2}{c}{zwabs$\times$power-law$+$zgauss} & $\chi ^{2}$/d.o.f. \\ \tableline
\Chandra/ACIS-I	& $r < 2\arcsec$	&	$kT$ [keV]	&	$4.89 ^{+6.07} _{-1.80}$	&	& &	\NH ~[$\times 10^{23}$\,cm$^{-2}$]	& $1.9 ^{+2.4} _{-1.2}$ &	45/42	\\
	&	& $Z$ [solar] &	$1.48 ^{+2.26} _{-1.13}$ & & & Photon index	& $0.71 ^{+0.82} _{-0.66}$ & \\ 
	&	&	Redshift	&	0.596 (fix)	&	&	& EW of \ion{Fe}{1} [eV]	& $134 ^{+207}_{-134}$	&	\\ \tableline
\Chandra/ACIS-I	& $2\arcsec < r < 3\arcmin $ &	&	&	$kT$ [keV] &	$10.9 ^{+1.8} _{-1.1}$	&	 &	&	89/122 \\
	&	&	&	&	 $Z$ [solar] &	$0.33 ^{+0.18} _{-0.16}$ &	&	&	\\
	&	&	&	&	 Redshift &	0.596 (fix) &	&	&	\\ \tableline
\Suzaku/XIS \& HXD and \Chandra/ACIS-I	& $r<3\arcmin$ &	$kT$ [keV] &	$2.95 ^{+0.53} _{-0.48}$ & $kT$ [keV] &	10.9 (fix) &	\NH ~[$\times 10^{23}$\,cm$^{-2}$]	&  $3.2 ^{+0.9} _{-0.8}$ &	790/732	\\
	&	&	$Z$ [solar] &	$0.76 ^{+0.63} _{-0.31}$ &  $Z$ [solar] & 0.33 (fix) & Photon index	& $1.54 ^{+0.27} _{-0.24}$ & \\ 
	&	&	Redshift	&	$0.599 ^{+0.004} _{-0.006}$	&	&	& EW of \ion{Fe}{1} [eV]	& $149 ^{+139} _{-58}$	&	\\
\tableline
\end{tabular}
\end{center}
\end{table*}

\section{Discussion}

With the simultaneous fit of the \Suzaku/XIS \& HXD and the \Chandra/ACIS-I data, we have discovered that the X-ray spectrum of the Phoenix cluster
can be approximated by three components;
two thin thermal plasma components with different temperatures and spatial distributions, and a power-law component in the cluster center.  
We also determine the redshift of the Phoenix cluster as  $z=0.599^{+0.004}_{-0.006}$  by the X-ray spectra alone. 
This value is consistent with the mean value ($z=0.596\pm 0.002$) of 26 member galaxies \citep{McDonald12}. 
In the following subsections, we separately discuss on the thermal emission from the ICM and the power-law component of the AGN.
We then discuss possible interactions between these components through the viewpoint of the feeding and feedback processes in cluster, BCG, and SMBH.
We refer all the physical parameters at the rest frame of $z=0.596$ unless otherwise specified.  

\subsection{Thermal emission from the ICM} \label{sec:ICM}

Although the ICM may consist of multi-temperature plasmas in pressure equilibrium,
it can be nicely approximated by two temperature plasma;
the low temperature plasma of $2.95 ^{+0.53} _{-0.48}$\,keV and the high temperature plasma of $10.9 ^{+1.8} _{-1.1}$\,keV.
The X-ray luminosity in the $2-10$\,keV band within 1.2\,Mpc are $L_{\rm X, low} = 1.0^{+0.4} _{-0.5} \times 10^{45}$\,ergs\,s$^{-1}$ 
and $L_{\rm X, high} = 7.0^{+0.7} _{-0.6} \times 10^{45}$\,ergs\,s$^{-1}$ for the low temperature and high temperature plasmas, respectively.
The high temperature plasma is prevailing in the whole cluster with the abundances of $0.33 ^{+0.18} _{-0.16}$\,solar, typical value for clusters. 
The low temperature plasma has a higher abundance of $0.76 ^{+0.63} _{-0.31}$\,solar and is confined in the core of the cluster.  
Possibly in the cluster core, a large amount of metals is supplied from supernovae in the BCG.
These features are commonly observed in nearby cool-core clusters with cD galaxies (i.e. BCG) at their centers \citep[e.g.][]{Matsushita11}.    

Following \cite{White97}, which is the same method as \cite{McDonald13}, we estimate the cooling rate to be $2290 ^{+1260} _{-770}$\, \MO \,yr$^{-1}$.
This value is consistent with that of \cite{McDonald13}.


\subsection{What is the origin of the \ion{Fe}{1} line?} \label{sec:origin}

As mentioned in section~\ref{sec:iron}, we discover the \ion{Fe}{1} line at the rest frame in the redshift of $z = 0.596$.  
We discuss whether the \ion{Fe}{1} line is due to the central AGN or to other possibilities.

One possibility is due to gain error of the \Suzaku/XIS.
The gain error of the \Suzaku/XIS is $\sim 0.1$\,\%\ at 6\,keV (6\,eV for the \ion{Fe}{1} line), 
which is estimated using several targets and \ion{Mn}{1} K${\alpha}$ lines from the calibration source of $^{55}$Fe \citep[e.g.][]{Koyama07b, Ota07, Ozawa09, Tamura11}.
The energy difference between \ion{Fe}{1} and \ion{Fe}{25} at the observer's frame is $\sim 200$\,eV, which is significantly larger than the gain error of the \Suzaku/XIS.

Second possibility is velocity-broadened (e.g. bulk motion and/or turbulence of the ICM) of \ion{Fe}{25} line (6.7\,keV at the rest frame), which may partly mimics as the \ion{Fe}{1} line. 
We fitted the X-ray spectra (the lower right panel of Figure~\ref{fig:XIS_HXD}) with a broadened line of the \ion{Fe}{25} line.
Then, we obtain the line width of \ion{Fe}{25} to be  $\sim 188$\,eV at the observer's frame, which corresponds to the velocity of $\sim 13000$\,km\,s$^{-1}$ at the rest frame \citep{Ota12}.
This velocity is too huge for the realistic motion of the ICM. 
Furthermore, \cite{McDonald12} reported that the X-ray surface brightness of the Phoenix cluster shows a relaxed morphology of the ICM, suggesting bulk motion is not prominent.

Third possibility is a contamination of other point sources than the AGN in the BCG within $3\arcmin$ radius.
We evaluate such an amount by examining the $2 - 10$\,keV intensity of the sources based on using the \Chandra \,11.9\,ks observation.
The emission within $2\arcsec$ from the central AGN contains total 501 counts in $2-10$\,keV band. 
Based on our fitting result shown in the lower left panel of Figure~\ref{fig:XIS_HXD}, 
we estimate that 361 counts is owing to the central AGN and the rest is primarily consists of the ICM thermal emission. 
On the other hand, there found six other point sources within $3\arcmin$ radius around the central AGN.  
The most intense one has 8 counts in $2-10$\,keV band, and sum of the counts from the six point sources is only 15 counts, which is 4\,\% of that of the central AGN.
Suppose that most of the six point sources are type I AGN with typical \ion{Fe}{1} EW of $\sim 100$\,eV 
(see e.g., the black triangles in the right panel of Figure~\ref{fig:Baldwin}), possible contribution for the \ion{Fe}{1} EW of the central AGN is only 4\,eV.

As discussed above, we can reject three possibilities for the origin of the \ion{Fe}{1} line.  
We hence regard the most natural idea is that the \ion{Fe}{1} line comes from the central AGN in the BCG of the Phoenix cluster.


\subsection{Properties of the type 2 QSO at the center} \label{sec:QSO}

We have discovered that the power-law component in the cluster center has a \ion{Fe}{1} K-shell line with EW of  $149^{+139}_{-58}$\,eV. 
The power-law photon index, $\Gamma$, is determined for the first time as  $1.54^{+0.27}_{-0.24}$.
The absorption column density of $3.2 ^{+0.9} _{-0.8} \times 10^{23}$\,cm$^{-2}$ is consistent with, but is more accurate compared with that in previous work \citep{McDonald12}.
The absorption corrected luminosity of $4.7 \pm 0.7 \times 10^{45}$\,ergs\,s$^{-1}$ ($2 - 10$\,keV) is slightly higher than that in \cite{McDonald12} 
but their consistency cannot be examined because they did not mention its error.

A large \NH, photon index in the range of $1.5-2.0$, and a strong \ion{Fe}{1} line are common features in type 2 AGNs \citep[][]{Awaki91}.  
Together with an extremely high X-ray luminosity, the central AGN of the Phoenix cluster can be regarded as a type 2 QSO.  
This is the second  case  of a type 2 QSO in a cluster after IRAS\,09104+4109 \citep[][]{Kleinmann88, OSullivan12}. 

The X-ray luminosity of $2.1 ^{+0.7}_{-0.8} \times 10^{46}$\,ergs\,s$^{-1}$  ($14-150$\,keV) is similar to the \Swift/BAT result of $1.4 \pm 0.9 \times10^{46}$\,ergs\,s$^{-1}$ within errors \citep{Cusumano10}.
As is seen in Figure 3, the power-law component (AGN) is dominated over the ICM plasma in the $14-150$\,keV band. 
Therefore the luminosity given by \cite{Cusumano10} is surely due to the type 2 QSO, 
and we see no large time variability in the type 2 QSO during the \Swift/BAT (2004 - 2010) and the \Suzaku \, observation (2010). 
Note that the luminosity estimated from the \Chandra/ACIS-I spectrum by us is $L_{\rm X} = 4.0 \pm 0.5 \times 10^{45}$\,ergs\,s$^{-1}$ in $2 - 10$\,keV
with the \NH\, and $\Gamma$ are fixed to the best-fit parameters of Table~\ref{tab:fit},
which is also consistent with the result of simultaneous fit.

Although the X-ray luminosity is exceptionally large as $10^{45}$\,ergs\,s$^{-1}$ ($2-10$\,keV),  
the \NH \,and EW are on the general trend of the correlation shown in \cite{Fukazawa11}, who compiled the \Suzaku \, results of 88 AGNs. 
The X-ray spectra of the type 1 and type 2 AGNs are interpreted in the unified scheme of AGNs. 
The differences in the observational properties of type 1 and type 2 AGNs are primarily due to observers' line of sight: face on (type 1) or edge on (type 2) to the molecular torus surrounding the nucleus.   
In this scheme, the X-ray spectra of type 2 AGNs consist of two components; one is penetrating through the torus, and the other is scattered at the surface of the torus. 
The former is called a direct component, and the latter is a reflection component. 
The \ion{Fe}{1} line is mainly associated with the reflection component. 
Since our spectral fit in section~\ref{sec:fit} implicitly assumed that the continuum flux of the AGN is dominated by the direct component only, 
we try the two-component structure of the type 2 QSO with the $pexmon$ model \citep{Nandra07} in XSPEC.
The $pexmon$ model represents an exponentially cut-off power-law spectrum reflected from neutral material. 
The direct, reflection components, and the fluorescent lines of Fe-K and Ni-K are included with a self-consistent manner.  
We assume the inclination angle, the cut-off energy of the power-law component, and the Fe abundance to be $\theta _{\rm i}= 60^{\circ}$, 300\,keV, and 1\,solar, respectively.  
Then, the X-ray spectra of the AGN component is reproduced with the best-fit value for the reflection fraction ($R$) of $0.77 ^{+0.48} _{-0.38}$, 
where $R$ is defined as the ratio of the solid angle of the reflector $\Omega$ to $2\pi$ steradian, i.e., $R=\Omega/2\pi$. 
This $R$ value and the initial assumption of viewing angle of  $\theta _{\rm i} = 60^{\circ}$ are consistent with the type 2 AGN picture of the torus edge-on geometry.

\citet{Ricci11} derived $R$ values for 165 Seyfert galaxies using the hard X-ray spectra with \textit{INTEGRAL}/IBIS \& ISGRI.  
They showed that Seyfert galaxies with $10^{23}$\,cm$^{-2} <$  \NH $< 10^{24}$\,cm$^{-2}$ (Compton thin AGNs), 
have $R$ of $2.2 ^{+4.5} _{-1.1}$ on average,  which are consistent with the type 2 QSO in the Phoenix cluster within large uncertainties. 
\cite{Ikeda09} performed  Monte Carlo simulations of the X-ray spectra in various torus geometries.  
The absorption column density and the EW of the \ion{Fe}{1} line for the type 2 QSO ($149 ^{+139} _{-58}$\,eV) in the Phoenix cluster 
are consistent with those 
from the simulation for the torus half-opening angle of 30$^{\circ}$ \citep[see the left panel of Figure~13 in][]{Ikeda09}.
Then, the expected EW of the \ion{Fe}{1} line is $\sim 200$\,eV.

\begin{figure*}[htbp] 
\begin{center}
\epsscale{1.0}
\includegraphics[width=72mm, height=54mm]{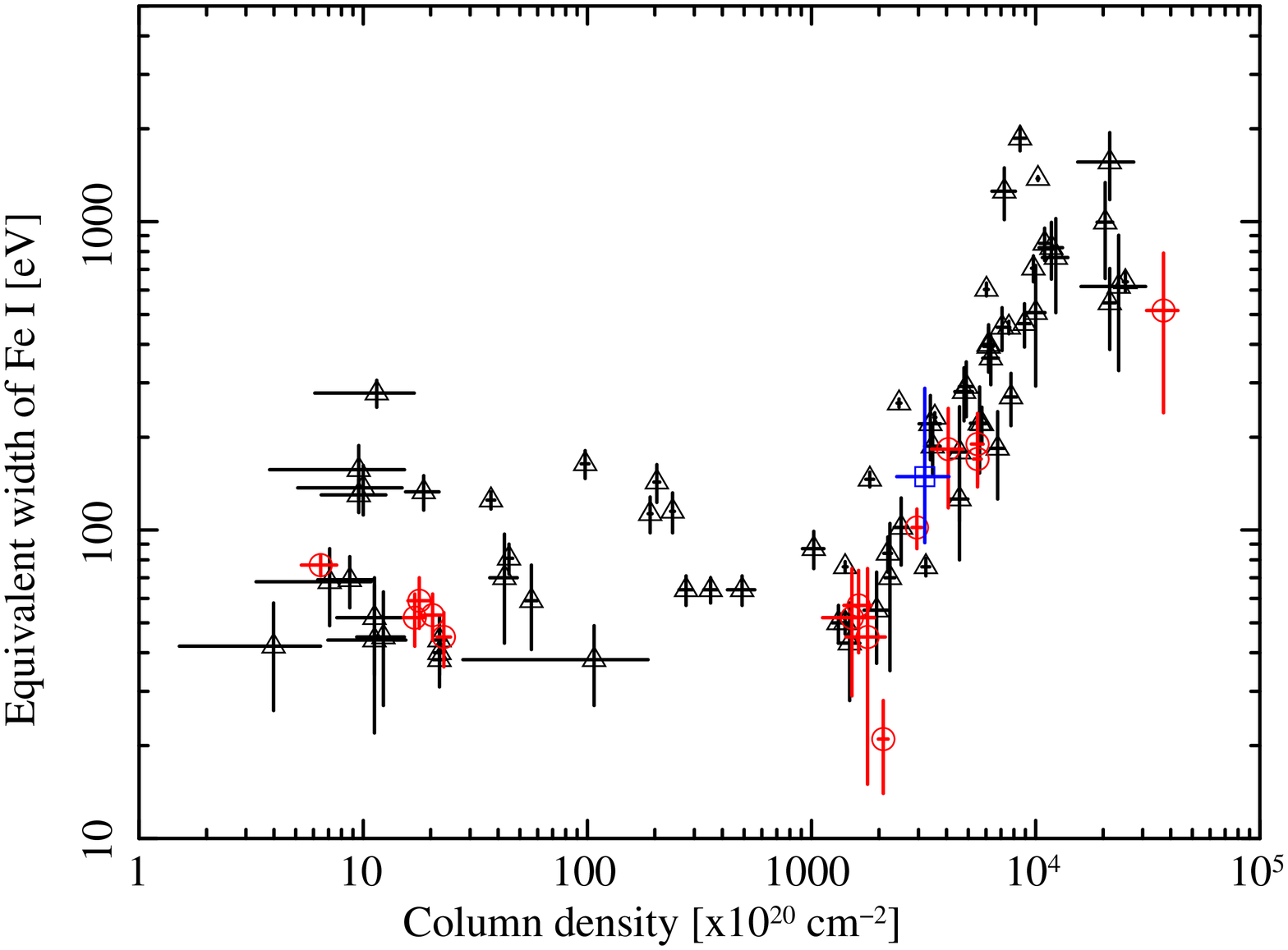}
\includegraphics[width=72mm, height=54mm]{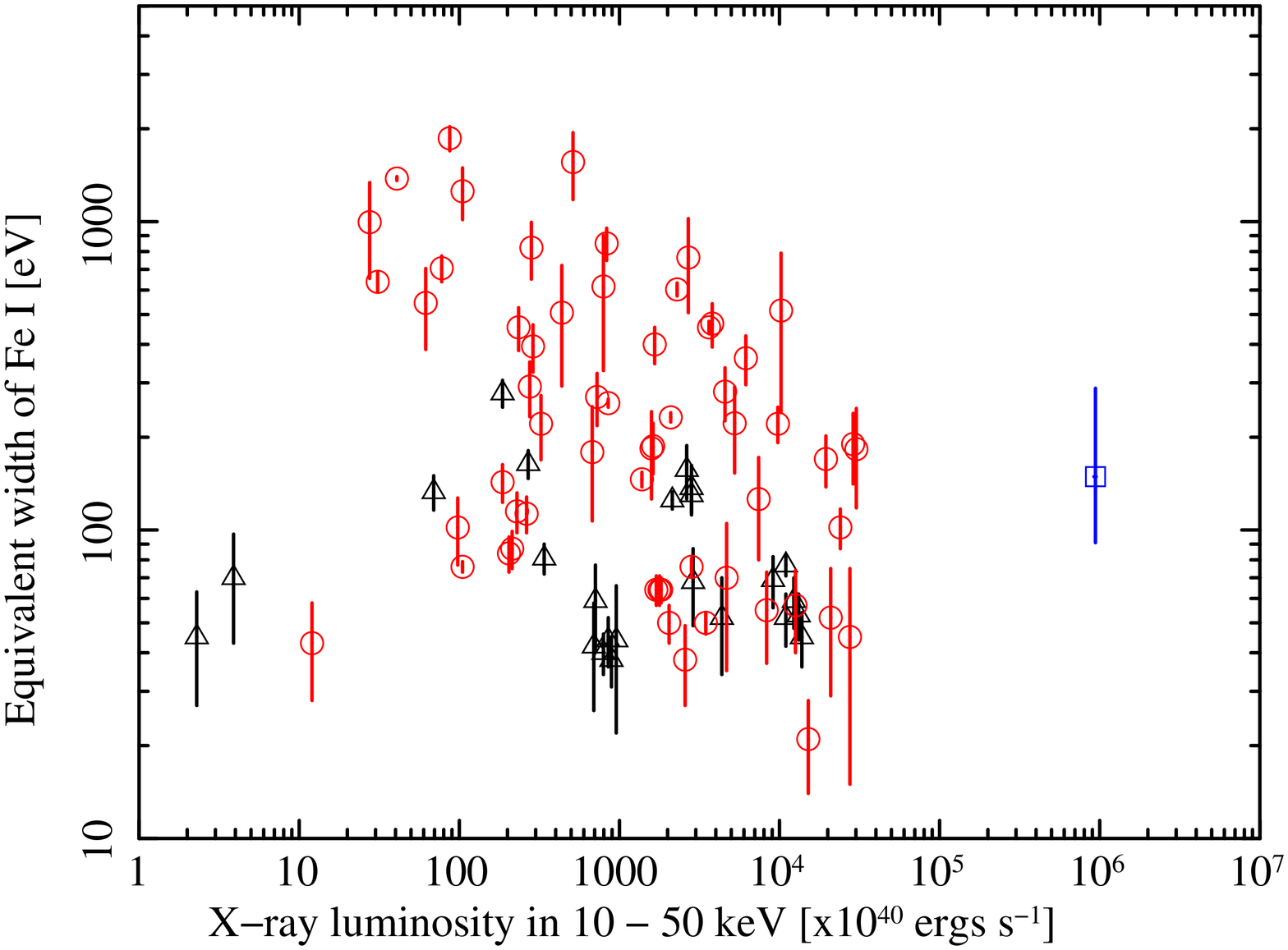}
\caption{
Left panel: relation of the EW of the \ion{Fe}{1} line and absorption column density (\NH).
The black triangles are AGNs with X-ray luminosity in $10 - 50$\,keV ($L_{\rm X, 10-50}$) of $L_{\rm X, 10-50} < 10^{44}$\,ergs\,s$^{-1}$ 
\citep[][but we excluded the AGN data of which no significant \NH, X-ray luminosity, and the EW of \ion{Fe}{1} were determined]{Fukazawa11},
while the red circles are $L_{\rm X, 10-50} > 10^{44}$\,ergs\,s$^{-1}$.
The blue square shows the type 2 QSO in the Phoenix cluster (this work).
Right panel: relation of the EW of the \ion{Fe}{1} line and the X-ray luminosity in $10- 50$\,keV.
The black triangles represent the AGNs with \NH \,$< 10^{22}$\,cm$^{-2}$, the red circles are \NH \, $> 10^{22}$\,cm$^{-2}$,
and the blue square is the type 2 QSO in the Phoenix cluster.
}
\label{fig:Baldwin}
\end{center}
\end{figure*}

The left panel of Figure~\ref{fig:Baldwin} shows a relation of the EW of the \ion{Fe}{1} line and the absorption column density (\NH) for our results of the Phoenix cluster and
86 Seyfert galaxies data \citep[][but we excluded the AGN data of which no significant \NH, X-ray luminosity, and the EW of \ion{Fe}{1} were determined]{Fukazawa11}.
The type 2 QSO in the Phoenix cluster is on the same trend of these type 1 and type 2 AGNs.
In general, a higher X-ray luminosity AGN exhibits a smaller EW of the \ion{Fe}{1} line, known as the X-ray Baldwin effect \citep{Iwasawa93}. 
The right panel of Figure~\ref{fig:Baldwin} shows the X-ray Baldwin effect by \cite{Fukazawa11} (the data selection is the same as the left panel of Figure~\ref{fig:Baldwin}).
The type 2 QSO in the Phoenix cluster has exceptionally larger flux than any other high luminosity AGNs.
Extrapolation of these data points by a linear function to the higher luminosity of $\sim 10^{46}$\,ergs\,s$^{-1}$,
gives the EW to be no larger than a few 10\,eV, far smaller than that of the type 2 QSO in the Phoenix cluster of EW $= 149^{+139} _{-58}$\,eV.
These indicate that the type 2 QSO has a torus of larger covering factor than those of the general trend of bright AGNs \citep[see Figure~5 of][]{Fukazawa11}.

\subsection{Type 2 QSO - ICM interaction} \label{sec:QSO-ICM}

In subsection \ref{sec:QSO},  
we show that the AGN in the BCG of the Phoenix cluster is a type 2 QSO with an unobscured X-ray luminosity of  $4.7 \pm 0.7 \times 10^{45}$\,ergs\,s$^{-1}$ ($2-10$\,keV). 
The EW of the \ion{Fe}{1} line of $\sim 150$\,eV is exceptionally large for the objects with such high luminosity.   
Furthermore, since the type 2 fraction is smaller for higher luminosity \citep[e.g.][]{YUeda03, Hasinger08}, type 2 QSO in the Phoenix cluster is a rare case. 
This may be related to other extraordinary properties of this object, i.e., a very active SMBH in the BCG, a huge cooling rate of the ICM, and a large star formation rate in the BCG.   

Assuming the bolometric correction factor of 130 \citep{Marconi04}, the bolometric luminosity of this type 2 QSO is $L_{\rm bol, QSO} = 6.2 \pm 0.9 \times 10^{47}$\,ergs\,s$^{-1}$, 
which corresponds to $\sim 27$\,\% of the Eddington limit.
Adopting the accretion efficiency of $\eta = 0.1$, the accretion rate is estimated to be $\sim 110$\,\MO \,yr$^{-1}$.  
This is $\sim 5$\,\% of the ICM cooling rate.
Other 35\,\% of the cooled gas may  be consumed by the violent star formation \citep[$\sim 798$\,\MO \,yr$^{-1}$,][]{McDonald13} in the BCG.
We denote these two rates as $\epsilon_{\rm acc}$ and $\epsilon_{\rm SFR}$.
The sum of $\epsilon_{\rm acc}$ and $\epsilon_{\rm SFR}$ (40\,\%) is larger than typical cool-core clusters.  

The mass of the torus can be estimated using  the observed column density of $3.2 \times 10^{23}$\,cm$^{-2}$  and assuming a spherical ring with the inner radius of 0.9\,pc. 
The inner radius of the torus is given by  the  $0.03 L_{43} ^{0.5}$\,pc relation \citep[][where $L_{43}$ is X-ray luminosity in unit of $10^{43}$\,ergs\,s$^{-1}$]{Suganuma06}.
The estimated torus mass is however largely dependent on the assumed outer radius,
such as  $3.0 \times 10^{4}$\,\MO \, $1.2 \times 10^{6}$\,\MO \ and $1.1 \times 10^{8}$\,\MO, and $1.1 \times 10^{10}$\,\MO \, for the outer radius of 1\,pc, 10\,pc, 100\,pc, and 1000\,pc, respectively. 
If the accretion rate of 110\,\MO \,yr$^{-1}$ has been constantly supplied by the torus, the torus mass is exhausted within $3 \times 10^{2} - 1 \times 10^{8}$\,yr,
shorter than the evolution time of a SMBH, BCG, and cluster.
Furthermore,  
if the outer radius is smaller than 1000\,pc, the lifetime of this torus is shorter than that of nominal QSO lifetime of $10 - 20$\,Myr reported by \cite{Hopkins05b}.
Possibly, some fraction of the cooling gas would be supplied continuously to the torus.
\cite{Taniguchi97} applied this idea to IRAS\,09104+4109 as a dust-enshrouded type 2 QSO in the center of a massive cooling-flow cluster \citep[e.g.][]{Fabian95, Crawford96, OSullivan12}
and estimated the mass of the torus to be $\sim 1 \times 10^{7}$\, \MO \,for a compact torus less than 10\,pc, or $\sim 1 \times 10^{9}$\, \MO \,for an extended torus of $\sim 100$\,pc.
\cite{Fabian90} shows that a QSO in the cluster center can be fueled in a self-sustaining way through Compton cooling of the surrounding the ICM.
In the type 2 QSO of the Phoenix cluster, the cooled gas would be also supplied continuously to the torus, and would finally accrete on the SMBH.

In idealized cool-core clusters, a large cooling rate should be converted to the cooling flow, which finally should appear as significant cold gas near at the cluster center. 
However no clear evidence for the fate of cold gas has been observed \citep[e.g.][]{Makishima01, Tamura01, Peterson01}. 
This is called as the "cooling flow problem" \citep[e.g.][]{Fabian94}.  
Some unknown mechanisms to suppress the cooling flow should be working. 
The most plausible explanation for the suppression of the cooling flow is heating by the AGN activity (called the "AGN feedback") \citep[e.g.][]{Fabian94, McNamara07}.
Its alternative is conduction of heat from the outer part of the ICM \citep[e.g.][]{Fabian03}.  

As we noted, the masses responsible to the violent star formation and accretion on the SMBH is 40\,\% of the cooling rate. 
The rest of 60\,\% of the gas may be cooled and deposited in the ICM within 100\,kpc to the BCG scale. 
The point here is that the fraction of 40\,\% is significantly higher than those of other nominal cool-core clusters \citep[$\sim 10$\,\% or less,][]{Blanton03, McDonald11, McDonald12}.  
A question is why the cooling flow suppression,
i.e., the AGN feedback,
in the Phoenix cluster is less effective than the other cool-core clusters. 
As one possibility, \cite{McDonald12} proposed that the Phoenix cluster is in a very rare epoch in the SMBH, BCG, and cluster evolutions, 
where the SMBH is powered by the cooling flow, but has not yet fully coupled with the ICM. 
Hence the quenching fraction of the total cooling is smaller than those in typical nearby cool-core clusters.  
If the AGN feedback is mainly due to jet interaction with the ICM \citep{Perucho11}, so called the kinetic-mode \citep{Fabian12}, 
the time from the latest onset of AGN activity
must be shorter than the light-crossing time of the cluster core, i.e., $\sim 0.3$\,Myrs. 
This is very short compared with typical timescale of AGN activity, and the Phoenix cluster must be in a very rare epoch. 

Alternatively, the AGN feedback to the ICM can take place through radiation from the QSO, called the radiative-mode or quasar-mode \citep{Fabian12}.
As mentioned in \cite{Fabian12}, this mode must be very important in the distant Universe but hard to be observed in nearby Universe. 
The Phoenix cluster is thus an exceptional
and possibly very important case.
On this point, we suggest that inefficient AGN feedback 
in the Phoenix cluster is related to the larger EW of the \ion{Fe}{1} line than that predicted from the general trend of the X-ray Baldwin effect.  
Possible explanation of the X-ray Baldwin effect is that strong X-rays from the central AGN would reduce the mass of the torus by the X-ray evaporation \citep{Pier95, Kallman04, Fukazawa11}, 
and hence reduce the EW of the \ion{Fe}{1} line.
As mentioned above, we suggest that some fractions of the massive cooling flow are supplied to the torus to compensate such an evaporation.
Then, at least, a significant part of the torus is maintained (i.e. not evaporated all neutral materials yet) against an intense irradiation of bright SMBH.
This torus may shield the radiation from the SMBH to suppress the heating of the ICM further.
Although we have few observational evidence of radiative-mode AGN feedback from a central AGN to ICM,
the shielding effect by the torus might be an important mechanism in SMBH and BCG evolution.


\section{Summary}

We have studied the X-ray spectra of the Phoenix cluster observed with the \Suzaku/XIS \& HXD and the \Chandra/ACIS-I, 
and have separately determined the ICM components and the central AGN component.
We confirmed that the ICM can be approximated by a low temperature ($kT = 2.95 ^{+0.53} _{-0.48}$\,keV) and high temperature ($kT = 10.9 ^{+1.8} _{-1.1}$\,keV) components. 
The low temperature component is concentrated at the cluster core and has a high abundance of $0.76 ^{+0.63} _{-0.31}$\,solar, 
while the high temperature component distributes over the cluster and has a abundance of $0.33 ^{+0.18} _{-0.16}$\,solar. 
These properties of the ICM are similar to those observed in nearby cool-core clusters. 
The major difference is its huge cooling rate of $\dot{M}_{\rm total} = 2290 ^{+1260} _{-770}$\,\MO \,yr$^{-1}$.

The X-ray spectrum of the central AGN in the Phoenix cluster is characterized with an strongly absorbed (\NH $= 3.2 ^{+0.9} _{-0.8} \times 10^{23}$\,cm$^{-2}$) power-law continuum 
plus the K-shell line from a neutral iron (\ion{Fe}{1}).  
The EW of the \ion{Fe}{1} line ($149 ^{+139} _{-58}$\,eV) and the absorption column density are typical for Compton-thin type 2 AGNs.  
However the EW is significantly larger than that of the general trend of the X-ray Baldwin effect, extrapolated to the luminosity as high as that of the type 2 QSO in the Phoenix cluster.

\acknowledgments
We are grateful to the anonymous referee for helpful suggestions and comments.
We thank all members of the \Suzaku \, \& \Chandra \, operation and calibration teams.
SU is supported by Japan Society for the Promotion of Science (JSPS) Research Fellowship for Young Scientist (A2411900).
This work is also supported by JSPS KAKENHI Grant Number 23340071 (KH), 24684019 (HN), and 23000004 (HT).

{\it Facilities:} \facility{\Suzaku ~(XIS, HXD)} \facility{\Chandra ~(ACIS-I)}.


\bibliographystyle{apj}
\bibliography{Type2QSO_accepted_arXiv_20130913}



\end{document}